\documentclass[prl,twocolumn]{revtex4}
\usepackage{epsfig}
\usepackage{epstopdf}
\usepackage{amsmath,amsfonts,paralist}
\usepackage{amssymb}

\usepackage{graphicx}
\usepackage{mathrsfs} 
\usepackage{color}
\usepackage{natbib,hyperref,url}

\definecolor{orange}{rgb}{0.9,0.7,0}

\newcommand{\BEQ}{\begin{equation}}
\newcommand{\EEQ}{\end{equation}}
\newcommand{\BEA}{\begin{eqnarray}}
\newcommand{\EEA}{\end{eqnarray}}
\renewcommand{\H}{{\cal {H}}}

\makeatletter
\newcommand\figcaption{\def\@captype{figure}\caption}
\makeatother

\begin{document}
\title{Imry-Ma criterion for long-range random field Ising model: short-/long-range equivalence in a field}
\date{\today}

\author{L. Leuzzi$^{1,2}$, G. Parisi$^{1,2,3}$}
\email{luca.leuzzi@cnr.it}
\affiliation{$^1$ Dipartimento di Fisica, Universit\`a {\em Sapienza},
  P.le Aldo Moro 2, I-00185 Roma, Italy. \\
  $^2$ IPCF-CNR, UOS Roma {\em Kerberos},  P.le Aldo Moro 2, I-00185 Roma, Italy.\\
  $^3$ INFN, Roma1,  P.le Aldo Moro 2, I-00185 Roma, Italy.}

\begin{abstract}
The Ising model in a random field and with power-law decaying ferromagnetic bonds is studied at zero temperature. Comparing the scaling of the energy contributions of the ferromagnetic domain wall flip and of the random field {\em \`a la} Imry-Ma we obtain a threshold value for the power $\rho$ of the long-range interaction, beyond which no critical behavior occurs. The critical threshold value is $\rho_c=3/2$, at a difference with the zero field model in which $\rho_c=2$. 
This prediction is confirmed by numerical computation of the ground states below, at, and above this threshold value. 
Some possible implications for the critical behavior of spin-glasses
in a field are conjectured.
\end{abstract}
\maketitle

{\em Introduction.}\qquad
According to the well known Imry-Ma argument \cite{Imry75,Chalker83,Fisher84} the  Random Field Ising Model (RFIM) 
with nearest-neighbor interaction does not
display any spontaneous magnetization in $D\leq 2$.
Spontaneous magnetization is, instead, present in $D=3$ where
a rigorous result has shown the occurrence of a finite dimensional  phase transition  \cite{Bricmont87,Bricmont88} and numerical
simulations \cite{Rieger93, dAuriac97, Sourlas99, Middleton02, Hartmann02} confirm this result. Further analysis by rigorous approach \cite{Imbrie84}, 
perturbation theory \cite{Villain85}, and RG transformations \cite{Aharony83, Berker84} have shown that $D=2$ is, actually, 
the lower critical dimension, and no evidence has been provided for the existence of any transition in dimension two,
 both for $T>0$ and at $T=0$. In the latter case  the relevant variable is the strength of the random magnetic field, i.e., 
 the square root of its variance, normalized to the ferromagnetic coupling.  Renormalization group arguments  show that the finite temperature transition is 
 dominated by the zero temperature fixed point.

In the present work we present the investigation of the  zero temperature critical behavior in a one-dimensional RFIM with long-range (LR) power-law decaying interaction.
Our aim is twofold: (a) to characterize the   threshold value  of the power 
above which the system does not undergo any phase transition; (b) to gain insight 
about the correspondence between LR models with a certain power of the interaction decay and  short-range (SR) models in a given dimension $D$ in presence of a field.
Our main result is that the critical threshold  value for the power corresponding
 to a lower critical dimension is $\rho_c=1.5$ in the 1D RFIM, rather than $\rho_c=2$ as in the
1D ferromagnetic model in absence of a field; the latter being the well known Kondo problem. \cite{Anderson70,Anderson71,Cardy81}. 
This has direct consequences on the determination of the lower critical dimension in presence of a field by means of the  analogy between long-range (LR) and short-range (SR) systems.

{\em SR$\leftrightarrow$LR connection with no field.} \qquad
We recall that a quantitative relationship can be established between the power-law $\rho$ of the LR interaction decay in a 1D lattice
and the dimension $D$ of a SR system displaying the same critical behavior. 
The requirement that the renormalized coupling constant has the same scaling dimension leads to 
\begin{equation}
\rho-1=\frac{2}{D}
\label{eq:rho_Da}
\end{equation}
Below the Upper Critical Dimension (UCD), though, i.e., for $\rho>\rho_{\rm mf}$ \footnote{$\rho_{\rm mf}=3/2$ in the ordered ferromagnet, whose UCD=$4$ and $4/3$ in the RFIM and in the spin-glass, where UCD=$6$.},
such relationship is not exact anymore. Moreover, it grossly fails at the Lower Critical Dimension (LCD), $D=1$ for the purely ferromagnetic model,
predicting a $\rho_c=3>2$ in a 1D  LR chain.
We can improve Eq. (\ref{eq:rho_Da}) by looking at  the behavior  of the renormalized space correlation function at criticality in SR model:
 $C(r)\sim r^{-D+2-\eta_{\rm sr}(D)}$. Requiring that at the 
  LCD the correlation function does not display any power-law critical decay, i.e.,  $D=2-\eta_{\rm sr}(D)$ and imposing
   the correct $\rho_c=2$, Eq. (\ref{eq:rho_Da}) is modified as
  \begin{equation}
{\rho-1}=\frac{2-\eta_{\rm sr}(D)}{D} 
\label{eq:rho_D}
\end{equation} 
By construction it is exact at the LCD.
The same relation holds for Heisenberg ferromagnets, where at the LCD ($D=2$), $\eta_{\rm sr}(D)=0$.
Eq. (\ref{eq:rho_D}) has been first obtained, in the framework of spin-glasses, 
by comparing the singular part of the free energy per spin in a LR system of $N=L^d$ spins and in a $D$-dimensional SR system with the same number of spins, $N=L^D$. The 
magnetic scaling exponents turn out to  follow the relationship $y_h^{\rm lr}=y_h^{\rm sr}(D)/D$, being
$2 y_h= D+2 -\eta$ \cite{Larson10,Banos12}.
Since in LR models, both with and without 
quenched disorder, the two point vertex function is not renormalized and $\eta_{\rm lr}=3-\rho$ \cite{Fisher72,Kotliar83, Leuzzi99} also in the infrared divergence regime, Eq. (\ref{eq:rho_D}) is recovered \cite{Larson10, Banos12, Leuzzi11, Ibanez12}. 

Eq. (\ref{eq:rho_D}) states that the critical behavior of the  two models, i.e., the $D$-dim. SR and the 1D LR models, should be similar for all $(\rho,D)$ couples between $(\rho_{\rm mf},UCD)$ and $(\rho_c,LCD)$ \footnote{Eq. 
(\ref{eq:rho_D}) has, actually, been tested both for spin-glasses \cite{Banos12} and ferromagnets \cite{Angelini13} finding 
discrepancies in both cases.}. For $\rho<\rho_{\rm mf}=3/2$ the system is in
the mean-field regime.  In the 1D Ising model without field this corresponds to 
$D>D_{\rm UCD}=4$.
As $D<D_{\rm UCD}$  infrared divergences occur in the vertex functions and a non-zero anomalous
exponent. In $D=3$ a good numerical estimate is $\eta=0.031(5)$ \cite{Pawley84}, corresponding to $\rho=1.656(2)$. 
In $D=2$, Onsager solution yields $\eta_{\rm sr}=1/4$ and the system is ``critically equivalent" to $\rho=15/8$ LR model.

By direct inspection, it is known that no transition is present at $\rho>\rho_c=2$. 
Exactly at $\rho=2$, though,  a phase transition does occur. This is the  Kondo transition in 1D magnetic chains \cite{Cardy81}.  On the contrary, the SR 1D Ising chain does not display any critical point. This is, actually, not unusual and it is due to a direct long interaction of iterfaces in LR models.
The critical behavior of the LR model at $\rho_c$ and of the SR model exactly at the LCD is often different: in some instances no transition is present in the SR model, while a transition may be present in the {corresponding} LR model. 
We anticipate that this is the case for the  RFIM as well: no transition at the SR LCD, but  a $T=0$ fixed point  with logarithmic scaling in the LR model at $\rho_c$.

We stress once again that the same Eq. (\ref{eq:rho_D}) holds for systems with quenched bond disorder, 
the so-called spin-glasses, in which
a rigorous result confirms $\rho_{c}=2$ \cite{Campanino87}.
Only the mean-field threshold value of $\rho$ is modified, because the relevant interaction term at criticality, and, thus, the upper critical dimension (UCD), is different: 
$\rho_{\rm mf}^{\rm sg}= 4/3$ \cite{Kotliar83,Leuzzi99}.

{\em SR$\leftrightarrow$LR connection in a field.} \qquad
As an external field is switched on a new  critical fixed point arises that is different from the zero-field fixed point. 
This is true both for systems with and without quenched bond disorder. 
Lower and upper critical dimensions appear not to decrease in all known cases. 
In particular, the critical dimensions of the RFIM increase to become
 $D_{UCD}=6$ and  $D_{LCD}=2$. 
The extension of Eq.  \ref{eq:rho_D} to the random magnetic case requires some care. 
{{Different definitions of the exponent $\eta_{\rm sr}$ are, indeed, 
 possible since connected and disconnected correlation functions decay differently and hyper-scaling does not hold \cite{Aharony76, Aharony83, Schwartz85,Bray85, Rieger93}. 
 Here, we  define an exponent $\bar \eta_{\rm sr}$ by the condition that the 
 Fourier transform of spin-spin {\em disconnected} correlation behaves in momentum space 
 as $k^{-4+\bar \eta_{\rm sr}}$, or equivalently in position space $C^{\rm disc}_{\rm sr}(r)\sim r^{-D+4-\bar \eta_{\rm sr}(D)}$, where the Schwartz-Soffer inequality holds:
 $\bar \eta_{\rm sr}\leq 2\eta_{\rm sr}$ \cite{Schwartz85}. The difference between   $2\eta$ and $\bar\eta$ decreases with the dimension \cite{Middleton02,Hartmann02},
 eventually tending to zero at the LCD. }}

We, now, present our study of the 1D LR RFIM. First, 
using an Imry-Ma-like argument we predict $\rho_c=1.5$. Further, we analyze  the critical behavior of the model at $\rho\sim \rho_c$
 by means  of numerical computations of the ground states properties at $T=0$ as function of the  strength of the  ferromagnetic interaction $J$.

{\em The long-range RFIM  and the Imry-Ma argument.}\qquad
The Hamiltonian of the LR 1D RFIM is 
\BEQ
\H=- \sum_{\langle ij\rangle}J_{ij}s_i s_j -\sum_i h_i s_i
\EEQ
where $J_{ij}=J |i-j|^{-\rho}$ and  $h_i$ is a random field with a bimodal distribution of zero average and variance $h^2$.

In a ordinary ferromagnet the cost to flip a domain of spins of length $L$ grows like $L^{2-\rho}$. 
As the random field is switched on 
this will compete with the energy of the orientation along the field going like $L^{1/2}$. 
According to the argument  developed by Imry and Ma for SR $D$-dimensional systems \cite{Imry75}, as $\rho>1.5$
 there will always be a size large enough for  the field  to destroy any ferromagnetic domain and no long-range order can be established.
The exponent value $\rho=1.5$ should, therefore, be the analogue of the LCD in nearest-neighbor interacting $D$-dimensional RFIM, i.e.  $D=2$ \cite{Imry75,Chalker83,Fisher84,Bricmont87,Bricmont88,Middleton02, Hartmann02}.

{\em L{\'e}vy lattice.}\qquad
In order to validate this analytic prediction we performed numerical estimates of the ground state properties at $T=0$ for the 1D RFIM on a 
L{\'e}vy lattice, that is a finite connectivity random graph equivalent to a fully connected LR model \cite{Leuzzi08}. In this
 dilute graph two sites $i$ and
$j$ are connected (i.e., $J_{ij}\neq 0$)
with a probability
\begin{equation}
P(J_{ij}=J) 
=\frac{|i-j|^{-\rho}}{\sum_r r^{-\rho}}
\end{equation}
where the sum runs over all possible distances realizable on the 1D chain of length $L$
and such that the total number of bonds is independent
from $\rho$ 
and equal to $z L$, 
where $z$ is the average spin connectivity.
For $\rho$ large enough one has a nearest-neighbor chain, whereas for $\rho=0$ the distribution of the connectivities is Poissonian and the system corresponds to an Erd\"os-R\'enyi graph.

{\em Numerical results.}\qquad
Using the  Minimum Cut algorithm \cite{Picard80,Hartmann98,Middleton02} of  the Lemon Graph Library \cite{lemon}  we have  computed thermodynamic observables on $T=0$ ground states of L\'evy graphs of different length, averaging over different realizations of random fields.
The computation has been performed varying the ferromagnetic coupling magnitude $J$.
 
 First, we present the numerical results at $\rho=\rho_c=1.5$ where we used  sizes ranging from $L=250$ to $L=256000$. The number of samples
 of disorder are $10000$ for $L\leq 64000$, $5000$ at $L=128000$ and $2000$ at $L=256000$. For each sample we compute the ground state
 for $41$ values of the $J$ coupling in the interval  $[0.2:0.4]$. The random field mean square displacement is kept constant, $h=1$. 
 
\begin{figure}[t!]
\center
\includegraphics[width=.99\columnwidth]{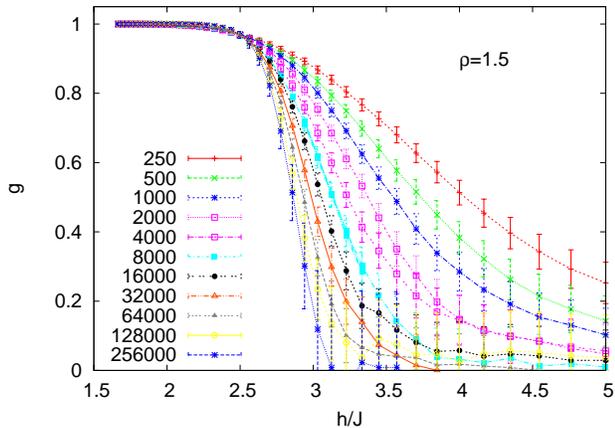}
\caption{Finite size Binder cumulants versus the strength of the random field in $J$ units at $\rho=1.5$. The critical value estimate by FSS analysis is $(h/J)_c=2.31(5)$. }
\label{fig:Binder_r15}
\end{figure}

\begin{figure}[t!]
\center
\includegraphics[width=.99\columnwidth]{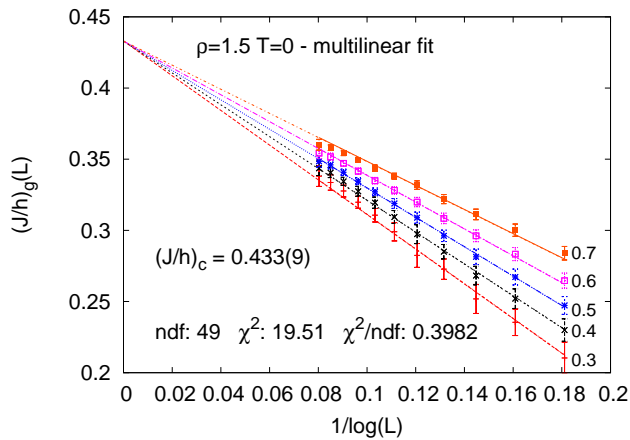}
\caption{Multilinear interpolation for the $J/h$ value at which finite size Binder cumulants significantly change value 
($g=0.3, 0.4, 0.5, 0.6$ and $0.7$) with $J_L(\ln L|g) = J_c + b_g/\ln L$. }
\label{fig:Hc_r15_FSS}
\end{figure}

To understand whether a critical behavior is there we study the finite size behavior of the Binder cumulant 
\BEQ
g=\frac{1}{2}\left(3-\frac{{\overline{\langle s\rangle^4}} }{{\overline{\langle s \rangle^2}}^2}\right)
\label{eq:Binder}
\EEQ
If, in the thermodynamic limit,  a phase transition occurs at a given critical field $h_c$, the Binder cumulant  will be one (long-range order) for $h<h_c$  and zero for $h>h_c$.
As $L$ increases, we observe that the various Binder curves tend to a limiting curve, cf. Fig. \ref{fig:Binder_r15},  
with a behavior that we will show compatible with a scaling logarithmic decay.
To estimate the critical point we look  at the values of $h/J$ that, for different sizes, yield the same $g$ value.
Specifically, in Fig. \ref{fig:Hc_r15_FSS} we show the behavior of the limiting inverse critical field value, $(J/h)_c$, as computed at every size for different fixed values of the Binder cumulant ($g=0.3,0.4,0.5,0.6,0.7$) versus $(\log L)^{-1}$. 
In the $L\to\infty$ limit all curves are compatible with a multiple linear fit in $1/\ln L$ yielding the estimate $(J/h)_c=0.433(9)$, or 
$(h/J)_c=2.31(5)$. 

\begin{figure}[t!]
\center
\includegraphics[width=.99\columnwidth]{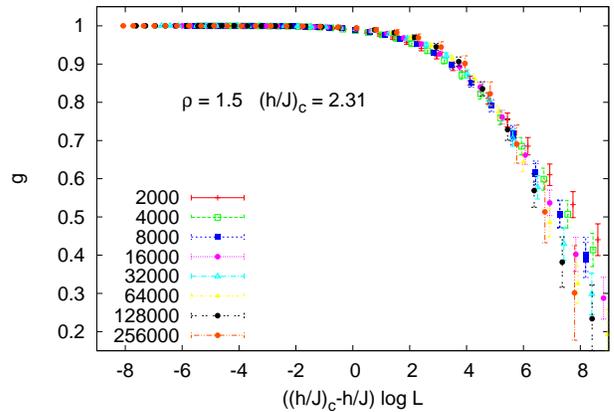}
\caption{Rescaled Binder cumulant versus $h/J$   at $\rho=1.5$.}
\label{fig:Binder_r15_scaling}
\end{figure}

\begin{figure}[t!]
\center
\includegraphics[width=.99\columnwidth]{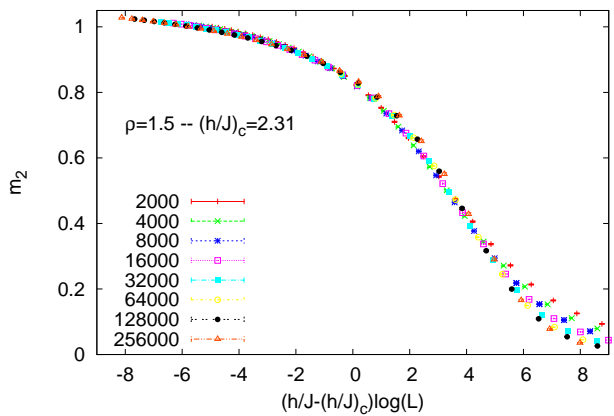}
\caption{Rescaled $m_2={\overline{\langle s\rangle^2}}$ curves versus $h/J$ at $\rho=1.5$. }
\label{fig:M2_h}
\end{figure}

In Fig. \ref{fig:Binder_r15_scaling} we plot the curves in the rescaled variable and observe a very good overlap in the critical region.
To further characterize the transition we look at the behavior of  magnetization momenta at the critical point.
In Fig. \ref{fig:M2_h} we, thus, present the behavior
 of the squared magnetization 
 around the estimated critical value
compatible with a logarithmic finite size scaling (FSS).

\begin{figure}[t!]
\center
\includegraphics[width=.99\columnwidth]{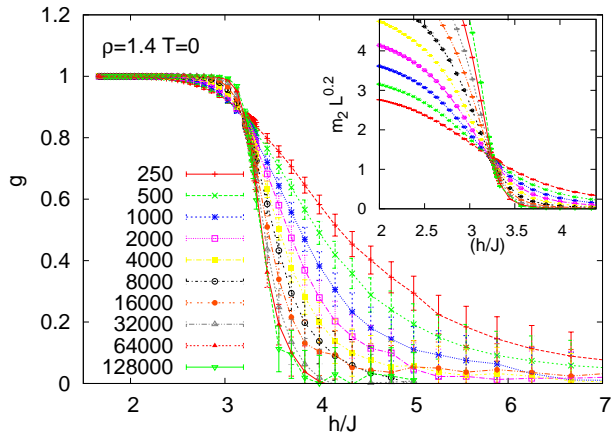}
\caption{Finite size Binder cumulants  at $\rho=1.4$ for $L=250, \ldots, 128000$. The critical field estimate is  $(h/J)_c=3.23(7)$. Inset: scale invariant $m_2 L^{0.2}  ={\overline{ \langle s\rangle^2}}L^{\bar\eta-3}$ vs. $h/J$, $\bar \eta=6-2\rho$. }
\label{fig:Binder_r14}
\end{figure}

As a comparison, we also present  the behavior of the Binder cumulant for values of the power $\rho$ slightly below and above $\rho_c$. 
For $\rho=1.4$ and $\rho=1.6$ we compute the ground states of systems of size between $L=250$ and $L=128000$
averaging over $10000$ disordered fields configurations on $51$ $J$ values.
At $\rho=1.4$, cf. Fig. \ref{fig:Binder_r14}, Binder curves cross each other at  finite $h/J$ and a FSS analysis 
of the crossing points yields a critical value $(h/J)_c=3.23(7)$. {{ Using the scaling property of the disconnected
correlation function ${\overline {\langle s\rangle^2}}\sim L^{3-\bar\eta_{\rm lr}}$, with $\bar\eta_{\rm lr}=3(2-\rho)$ (this formula should hold for  $\rho\in[1,3/2]$,
 from the FSS of the crossing points (cf. inset of  Fig. \ref{fig:Binder_r14}) we further obtain the estimate $(h/J)_c=3.266(2)$
 and from the FSS of the derivatives of ${\overline{ \langle s\rangle^2}}$ we estimate $1/\nu=0.316(9)$.}}

\begin{figure}[t!]
\center
\includegraphics[width=.99\columnwidth]{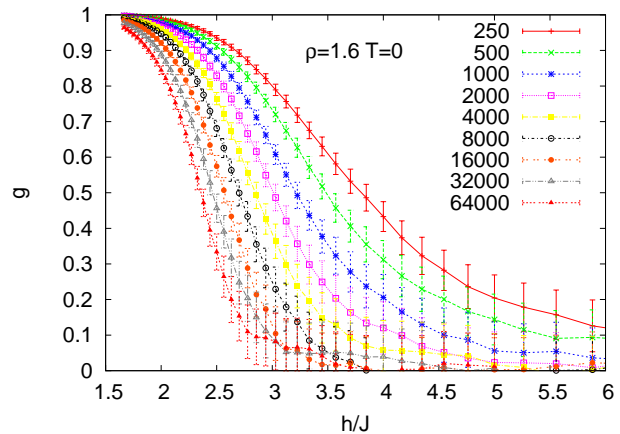}
\caption{Finite size Binder cumulants  at $\rho=1.6$ for $L=250, \ldots, 64000$.}
\label{fig:Binder_r16}
\end{figure}

For $\rho=1.6$, on the contrary, no crossing is observable and the non-zero Binder values continuously run away towards larger  and larger fields, cf. Fig. \ref{fig:Binder_r16},
 compatibly with the claim of  absence of  transition  above $\rho=1.5$.

 Eq. (\ref{eq:rho_D}) is a not too bad approximation for what concerns  the 
transition without field.
With no field, at $\rho=3/2$, Eq. (\ref{eq:rho_D}) would predict a mean-field transition  (corresponding to $D=4$); non-mean-field transitions would be expected 
at $\rho=1.6545$  (corresponding to $D=3$, with $\eta_{\rm sr}(3)=0.0364(5)$  \cite{Pelissetto02}),  
and $\rho=1.875$ (corresponding to $D=2$, 
$\eta_{\rm sr}(2)=1/4$ \cite{Onsager43}); eventually, the``LCD"-equivalent exponent value 
corresponding to $D=1$ ($\eta_{\rm sr}(1)=1$) woud be $\rho_c=2$.

Such predictions have been recently numerically investigated showing that  the critical exponents at $\rho=1.6546$ and $1.875$ do not strictly correspond to, respectively, $2D$ and $3D$ critical exponents  \cite{Angelini13}: if for 3D, nearer to the mean-field threshold, numerical estimates are still consistent with each other, in 2D
 they appear not compatible anymore. 
 A similar trend has been identified in spin-glasses where LR systems with values of $\rho$ equivalent to 4D and 3D have been analyzed \cite{Banos12}.

When the random field is switched on and a new fixed point for the RG flow arises the situation changes. 
The mean-field threshold is now $\rho_{\rm mf}=4/3$ (UCD=$6$). The Imry-Ma argument and the simulations presented in this work clearly show that the 
threshold for the critical behavior is $\rho_c^h=1.5$.

{\em Discussion and conclusions.}\qquad
{{From the present work we clearly understand that 
the reference values for $\rho$ are different from those obtained in absence of a field.
We obtain such evidence by means of an Imry-Ma-like argument and a numerical study of the zero temperature ground states of 
the RFIM on a L\'evy lattice for system sizes ranging fro $250$ to $256000$ spins. 
Specifically, we find that in the 1D 
RFIM with LR interactions no transition is present for $\rho>1.5$  and that at $\rho=1.5$ a $T=0$ fixed point is still present
with a logarithmic scaling.}}

{{In presence of a random field we can reformulate the ``$\rho-D$" relationship Eq. (\ref{eq:rho_D})  in terms of the 
anomalous exponent $\bar \eta_{\rm sr}(D)$, rather than $\eta_{\rm sr}(D)$.
That is, we consider the most divergent correlation function at criticality in a SR system in dimension $D$: the disconnected one.
 At the lower critical dimension ($D=2$), where 
$D-4+\bar \eta_{\rm sr}(D)=0$, the
 threshold value of the power $\rho$ has to be equal to the maximum one compatible with the existence of a transition: $\rho_c=3/2$.
This leads to
\begin{equation}
\rho-1=\frac{2-\bar\eta_{\rm sr}(D)/2}{D} 
\label{eq:rho_Drandom}
\end{equation}
 yielding the value of $\rho$ corresponding to a SR model in $D$ dimensions. As Eq. (\ref{eq:rho_D}) in zero field, 
Eq. (\ref{eq:rho_Drandom}) is exact, at all events, at $D=$UCD and LCD.  Since in the latter case $\bar \eta_{\rm sr}\simeq 2\eta_{\rm sr}$ we notice that in this particular case 
 Eq. ({\ref{eq:rho_Drandom})
  coincides with Eq. (\ref{eq:rho_D}). For both Eqs. (\ref{eq:rho_D})  and (\ref{eq:rho_Drandom}) the LCD equivalent value of $\rho$ is the correct one: $\rho_c=1.5$.}}
{{It is important to stress that a given value of $\rho$ corresponds to completely different critical behaviors and to different
dimensions of short-range critically equivalent systems if the field is present or absent.
 As an instance $\rho=1.5$ is the mean-field threshold  in the Ising ferromagnetic model, corresponding  to UCD $D=4$
and it is the critical threshold in the RFIM, corresponding to LCD $D=2$.  
}}

Does this relationship hold also in presence of {\em random bonds}, besides {\em random fields} ?
The Imry-Ma argument is specific for the RFIM and cannot be exported to spin-glasses because
these more complicated systems lack any long-range order in the frozen phase. 
Therefore, a quantitative estimate of the threshold value $\rho_c^h$ corresponding to the SR LCD is beyond the reach of the analysis presented here.
Contrarily to the ordered bond model,  in spin-glasses  the UCD does not increase by applying a random field
($\rho_{\rm mf}=4/3$). Nor the LCD, that
remains equal to $D=2.5$ according to a computation of interface free energy \cite{Franz95}.

In LR systems in a field, though, we would be surprised to observe a spin-glass phase
 for values of $\rho>\rho_c^h=3/2$ for which no ferromagnetic transition is present in absence of bond disorder.
This suggests that much  caution should be taken in numerical data interpretation in LR spin-glass systems
 in presence  of a field. Above all when
 ultrametricity \cite{Katzgraber12} or lack of Almeida-Thouless transition \cite{Katzgraber05} are
 tested at $\rho>1.5$.

{\em Acknowledgments}\qquad
The authors thank Maria Grazia Angelini, Victor Martin Mayor, Federico Ricci-Tersenghi and David Ylanes
  for interesting discussions on the topic of the work and careful readings on the manuscript.
The research leading to these results has received funding from the European Research Council
(ERC) grant agreement No. 247328,
from  the People Programme (Marie Curie Actions) of the European Union's Seventh Framework Programme FP7/2007-2013/ under REA grant agreement No. 290038  (NETADIS project) and 
from the Italian MIUR under the Basic
Research Investigation Fund FIRB2008 program, grant
No. RBFR08M3P4, and under the PRIN2010 program, grant code 2010HXAW77-008.

\end{document}